# Strategic Governance of AI Models in Earth Science

Makoto Kelp*[1], Amirhossein Arzani[1], Patricia Castellanos[2], Paul Griffiths[3], Ivan Higuera-Mendieta[4], Manuel Perez-Carrasco[5], Viral Shah[6], Patrick Obin Sturm[7], James Weber[8]

[1]University of Utah
[2]NASA Goddard Space Flight Center
[3]University of Bristol
[4]Stanford University
[5]Center for Astrophysics | Harvard & Smithsonian
[6]GESTAR II, NASA Global Modeling and Assimilation Office
[7]University of Southern California
[8]University of Reading

*Co-authors are listed alphabetically by last name*

*****Corresponding author: Makoto Kelp
Email: makoto.kelp@utah.edu

**Executive Summary:**
AI foundation models pretrained on weather and climate data are increasingly fine-tuned to Earth science tasks well beyond weather forecasting. Their development and adoption are outpacing the scientific community's ability to evaluate them. These models are judged almost entirely by benchmark skill metrics, which measure how closely a forecast reproduces a reference product but not whether a model represents the physical processes governing the system it predicts. Forecast skill and physical reliability are therefore distinct properties. The distinction is most consequential under the nonstationary conditions of a changing climate for which these models were never trained. We identify five priorities for the physical evaluation of AI models in Earth science from task-specific emulators to foundation models, spanning training data, fine-tuning, behavioral testing, mechanistic interpretability, and output validation. We recommend three activities for the coming decade: 1) open AI-ready evaluation datasets, 2) a shared reporting standard for physics-based evaluation, and 3) a dedicated research program on the safety of these models.

## 1. Issues and Significance

The governance of AI models is an urgent need for NASA Earth Science as these models are being developed and adopted faster than the scientific community can evaluate them. The concern spans the full range of AI models now used in Earth science from task-specific emulators to large foundation models. However, the risk is most acute for foundation models whose rapid, low-cost adoption magnify every evaluation gap described herein. Earth system foundation models are large neural networks pretrained using self-supervision on diverse datasets to perform tasks such as atmospheric and ocean phenomena prediction (Bommasani et al., 2022; Zhu et al., 2026). This white paper focuses on foundation models pretrained on weather and climate reanalysis data, which are some of the most developed and most widely adapted (S. Chen et al., 2023; Mukkavilli et al., 2023), though the evaluations described here apply to AI models in Earth science broadly. These models can be adapted, or "fine-tuned", to more specific tasks through additional training on smaller, specialized datasets (Zhang et al., 2025). Earth science foundation models have improved quickly both in accuracy and the range of tasks they can perform. NASA-IBM's Prithvi WxC (Schmude et al., 2024) and Microsoft's Aurora (Bodnar et al., 2025), alongside task-specific AI weather models such as Google's GraphCast (Lam et al., 2023) and ECMWF's AIFS (Lang et al., 2024), now produce global weather forecasts with skill that matches or exceeds conventional numerical weather prediction at a fraction of the computational cost (Bi et al., 2023, 2022; L. Chen et al., 2023; Pathak et al., 2022; Vaughan et al., 2024). Because a single pretrained model can be fine-tuned to many downstream tasks, these AI models are increasingly adapted well beyond weather forecasting. Examples include atmospheric composition and tropical cyclone prediction with Aurora, and gravity wave parameterization and localized renewable energy forecasting with Prithvi WxC. Many of these tasks depend on processes the underlying model was never trained to represent, and it is unclear whether a model whose internal representations carry a strong inductive bias toward weather transfers reliably to physically disparate tasks or out-of-distribution events. However, models are increasingly fine-tuned to new tasks regardless of whether they are suited to them.

AI models cannot be validated in the same way conventional process-based models can. A numerical weather, climate, or chemical transport model is built from explicit equations and conservation laws that govern its behavior, so its errors can be traced to specific physical assumptions (Bey et al., 2001; Keller et al., 2021; Molod et al., 2015). AI models are not created in this way. Their model architecture is fixed, but the internal representations that form during training are not easily or explicitly specified, cannot be controlled directly, and are not required to correspond to any physical process (National Academies of Sciences, Engineering, and Medicine, 2025). Lacking a physical basis to inspect, the scientific community has come to judge these AI models largely through benchmark performance (Kasteleyn et al., 2026; Sturm and Wexler, 2022; Ullrich et al., 2025). The field's reliance on benchmark leaderboards compounds this, rewarding models that optimize for a score rather than models whose internal representations of the physical system are valid (Corley et al., 2026). Root mean square error (RMSE) and correlation against reanalysis data permit direct comparison between AI models and have driven rapid progress in AI weather prediction (Rasp et al., 2024, 2020). But these metrics reward agreement with a reference product and reveal little about whether a model has captured the processes that produced the original data, or, if it captures them, whether it also inherits the reference product's weaknesses in parametrization and resolution. A model can achieve low error by learning associations in the training data without representing the mechanisms those associations reflect, and output metrics cannot distinguish the two cases. Such a forecast also yields little understanding that could be transferred back to improve conventional modeling systems or to inform users about the inherent limits or risks in using these outputs for decision making.

This lack of transparency becomes critical under nonstationary and out-of-distribution conditions. An AI model trained to reproduce reanalysis datasets has no guarantee of accuracy once conditions move beyond it nor does it have statistical guarantees when conditions are within its training scope (Angelopoulos et al., 2023), yet prediction under such conditions is central to a changing climate. The concern is amplified under extreme heat, wildfire smoke, and shifting emissions, land use, and ecology, conditions poorly represented in the historical record and dependent on relationships the model never encountered during pretraining (IPCC, 2023). Predicting rare events is a significant challenge for data-driven models and might require specialized development (Sapsis, 2021). These are also the conditions under which prediction is most consequential for society. Therefore, forecast skill and physical reliability are distinct properties and existing benchmarks assess only skill (Husain et al., 2025). We argue that a model's representation of the underlying processes should itself be a criterion of evaluation, particularly for the conditions under which these models will increasingly be used (e.g., climate projections) and for which they were never trained.

These technical concerns are made more pressing by how AI models, and foundation models in particular, are being adopted (Gal and Casper, 2025; Nuruzzaman Nobel and Tan, 2026). Fine-tuning demands far less computation, data, and expertise than developing a conventional modeling system, so adoption is projected to grow rapidly across research institutions, operational forecasting centers, and private companies (Hickman et al., 2025). That same low cost favors rapid deployment and the groups that build and fine-tune these models may face little pressure to characterize physical behavior when benchmark skill alone satisfies current standards of publication and use. This growth in adoption also coincides with a broader shift toward the privatization of Earth data and mounting pressure on the public institutions that provide open observations and forecasts (Soranno and Cheruvelil, 2025). As a result, the open data underpinning independent evaluation is becoming less secure. The same concern extends to the AI models themselves. Independent physical evaluation requires access to model weights and training details that closed systems do not provide. Open models are therefore a precondition for verifying that a prediction can be trusted. As fine-tuned models enter operational forecasting and decision support, unexamined and potentially non-physical behavior translates directly into risks to public health and resource management. Closing this gap is a technical, scientific, and stewardship issue, as well as an AI safety concern.

## 2. Guiding Questions for Evaluating AI Models in Earth Science

The rapid rise of AI models in Earth science, and the projected proliferation of foundation models, raises four fundamental questions that the NASA Earth Science community should prioritize:

*1. Can we determine whether an AI model predicts through physical mechanisms rather than statistical correlations?*

*2. Will a model that is accurate on the historical record remain reliable under the nonstationary conditions for which it was never trained?*

*3. How do pretraining and fine-tuning choices determine a model's skill and physical reliability?*

*4. What standards, open observations, and governance would make physical reliability a condition of use?*

The first three questions concern the safety and trustworthiness of these models and are addressed by the evaluation priorities in Section 3. The fourth concerns their governance and is addressed by the recommended activities in Section 4.

## 3. Priorities for Evaluating and Understanding AI Models

Whether a pretrained or fine-tuned model is physically consistent cannot be diagnosed simply from its agreement with existing benchmark datasets. An AI model can attain low error by learning quantities that covary with its target rather than the processes that govern it (Silva and Keller, 2024). For example, a model predicting surface ozone may succeed largely by learning its correlation with temperature and then fail when emissions or wildfire smoke shift that correlation (Schnell and Prather, 2017). Interpretability of Earth science AI models is only beginning but the first studies already show that a skillful model can rely on structure unrelated to the physical processes it appears to predict. In GraphCast, sparse autoencoders have recovered internal features that activate with tropical cyclones, atmospheric rivers, and large-scale precipitation, alongside other features that encode the model grid and geography rather than any physical process (MacMillan and Ouellette, 2025). In the Aurora Earth system foundation model fine-tuned for atmospheric chemistry, individual pollutants are forecast skillfully at lead times up to five days (Bodnar et al., 2025). The model nonetheless fails to preserve the photochemical relationship linking nitrogen oxides ($NO_x$) and ozone, which it was never trained on explicitly but had the inputs to learn (Hu et al., 2026; Figure 1). Physical behavior therefore differs between models and even between variables within a single model. Whether that behavior is physically consistent cannot be inferred from skill and determining it requires examining how a model represents the system it predicts. Methods to do so exist in the interpretable machine learning literature for weather (McGovern et al., 2019; Yang et al., 2024) and more broadly in the practice of evaluating large AI systems where behavioral testing and data curation are used to establish whether a model can be trusted beyond the conditions on which it was tested (Hendrycks et al., 2022; Olah et al., 2020). However, their application to Earth science AI models has barely begun. We propose developing that application through the five priorities below, illustrated for a foundation model in Figure 2.

**Training data filtering.** The behavior of an AI model is determined first by the data on which it is trained, yet this influence is rarely examined. These models are typically trained on whichever large datasets are available, rather than on data chosen to test how inputs affect what a model learns. The dominant approach pretrains on reanalysis datasets, which are attractive for their global coverage and internal consistency. Reanalysis assimilates observations into a physically consistent gridded state well-suited to weather prediction, but it also smooths the atmosphere in ways that can suppress the sharp gradients and extremes a model must capture (Abel et al., 2026; Hersbach et al., 2020). For atmospheric composition, it further blends emissions, chemistry, transport, and meteorology into a single product that obscures the individual processes a model would need to learn. The choice of training data raises two distinct questions. The first concerns which data sources, variables, and resolutions produce the most skillful forecasts. Training directly from observations rather than reanalysis is one active alternative, though such systems do not yet match the skill of reanalysis-trained models (Lean et al., 2026). Training on the output of process-based models is another, which exposes a model to physically consistent fields but risks inheriting those models' biases and missing processes. Indeed, the likely importance of training on the output of process-based models means that the continued development of such models remains critically important. The second question concerns which data best allow a model to learn identifiable physical processes rather than correlations. Combining observations, reanalysis, and process-resolved model output within a multi-fidelity AI setting may serve this goal and could in turn reveal gaps in the physics-based models that such output comes from. Neither question has been studied systematically for AI models in Earth Science.

Training data attribution (Cheng et al., 2025) offers a path to study both questions. These methods quantify the influence of individual training examples on a model's predictions and can now scale to models with billions of parameters through influence functions (Grosse et al., 2023; Ilyas et al., 2022; Koh and Liang, 2017) and gradient-based estimators such as TRAK (Park et al., 2023). Applied to an Earth system AI model, attribution would identify which segments of the training record a given forecast depends on, revealing whether a prediction of an extreme event traces back to physically similar events or to unrelated correlations in the reanalysis. This practice would turn training data curation from a typical documentation exercise into a causal diagnostic, and it connects this data priority directly to the interpretability priority below. Documenting the composition and processing of training data as standard practice and characterizing how these choices affect physical behavior are among the least expensive means of understanding and improving these AI models.

**Fine-tuning best practices.** How a pretrained model is adapted determines whether it behaves correctly on a task for which it was not designed. A range of methods are now in use, including full supervised fine-tuning, parameter-efficient approaches such as low-rank adaptation (LoRA), and adapter layers (Hu et al., 2021; Zhang et al., 2025). It is not known how fine-tuning preserves or degrades the physical behavior a model acquired during pretraining (Kumar et al., 2022; Shuttleworth et al., 2025), although a balance between retaining prior information (i.e., avoiding catastrophic forgetting) and efficiently learning new information from the fine-tuning data must be struck (Elhadidy et al., 2026). This concern is greatest when the target depends on processes that pretraining did not represent. For example, a model pretrained on weather contains no representation of biological or biogeochemical processes, so a version fine-tuned to predict ocean primary productivity or land carbon flux must infer them entirely from correlated physical fields. Studying how fine-tuning changes a model also illuminates how these models acquire the ability to represent a process. Methods that remove learned behavior, such as machine unlearning (Baiman et al., 2026; Bourtoule et al., 2021), offer a complementary way to probe what a model has internalized. Systematic study of how fine-tuning choices determine skill and physical behavior is therefore a priority commensurate with the central role of fine-tuning in how these models are used.

**Testing consistency under extremes and out-of-distribution conditions.** Extreme and record-breaking events are among the most consequential conditions to forecast and a common failure mode of AI models. AI models have been shown to underperform physics-based models on record-breaking extremes (Zhang et al., 2026). Standard benchmark metrics reward reproduction of the training record and cannot reveal how a model behaves beyond it. The evaluation protocol itself compounds the problem. That is, Earth system data are spatially and temporally autocorrelated, so random train-test splits place test samples near training samples and overstate generalization (Kattenborn et al., 2022; Rolf et al., 2024). Given the nonstationarity of the Earth system, spatially and temporally aware evaluation should be standard practice for AI models and their fine-tuned versions. Controlled perturbation experiments address this gap by altering a specific input variable and examining whether the resulting forecast changes in a physically plausible way, which tests the sensitivities a model has learned rather than its agreement with a reference (Hakim and Masanam, 2024). Increasing an emissions input and testing whether the corresponding pollutant rises is one such experiment. Raising a temperature field and testing whether the dependent variables respond as thermodynamics dictate is another. Each distinguishes a learned physical dependence from a potentially coincidental correlation. Such idealized perturbation experiments mirror those used to probe conventional physical models, and while they cannot prove a model uses physical mechanisms, they test whether its responses are physically plausible. These tests require the model to take the relevant variable as an input, which many models do not. Even then a physically correct response is not guaranteed, because the model holds no explicit representation of the governing process. Perturbation and out-of-distribution testing is established in the evaluation of large machine learning systems (Feng et al., 2026; Liu et al., 2024) and should become standard for Earth system AI models. Additionally, the precise definition of “out-of-distribution condition” might not be obvious in complex settings once the model is deployed and developing automated out-of-distribution inference safeguards is another open problem.

**Mechanistic interpretability of model internals.** The priorities above assess a model by its inputs and outputs. Mechanistic interpretability instead looks inside the model at the internal representations that generate a forecast to determine how that forecast is produced. Existing explainable AI (XAI) methods, such as saliency mapping, layer-wise relevance propagation, and Shapley additive explanations (SHAP), identify which input variables influence a forecast (Bommer et al., 2024). They are largely correlational and an output can be attributed to the correct inputs while the internal computation that produced it bears no relation to the underlying physics (Silva and Keller, 2024). Determining whether a model represents the processes it depends on therefore requires examining not what enters and leaves the model, but the computation in between (which post-hoc XAI methods do not reach). We argue that the field should move beyond input attribution toward mechanistic interpretability. Developed for large language models, mechanistic interpretability reverse-engineers trained neural networks to identify the internal features a model represents and the computational pathways that connect them (Bereska and Gavves, 2024; Nanda et al., 2023; Rai et al., 2025). Alternatively, one can build inherently mechanistic models through by-design interpretable approaches such as analytical model discovery (Brunton et al., 2016; Viknesh et al., 2026). These features are tested causally by intervening on an internal representation and measuring the resulting change in the model's output (Hatfield-Dodds, 2023; Olah et al., 2020). Applied to Earth AI models through sparse autoencoders and causal feature ablation, these methods can identify the features a model uses to make a prediction and determine whether they correspond to the physical processes that govern the target or to correlations that will not generalize (Hu et al., 2026; MacMillan and Ouellette, 2025). We regard this as a central priority because the causal nature of these tests begins to trace the internal pathways by which a model transforms its inputs into a forecast. Such an account is closer in kind to the explicit equations of a conventional modeling system and is particularly suited to anticipating a model's behavior outside its training distribution.

**Post-hoc validation over 'physics-informed' learning.** Two general approaches aim to keep AI model forecasts physically consistent, but they differ in where that consistency is imposed. Physics-informed machine learning builds physical information into the model during training, usually through modified loss functions or architectural design (Karniadakis et al., 2021; Kelp et al., 2020; Sturm and Wexler, 2022). On the other hand, post-hoc validation checks a model's outputs against physical laws after training. For large models, we argue that the second is the higher priority (Kasteleyn et al., 2026). Physics-informed machine learning methods have proven effective for problems with well-specified governing equations (Raissi et al., 2019), but the atmosphere and broader Earth system are governed by processes that are numerous, incompletely known, and imperfectly parameterized, so the constraints available to impose are necessarily partial or impractical to implement. Enforcing only the known relationships adds strong inductive biases that can limit an AI model's capacity to learn the novel or unexpected structure in data that gives data-driven models much of their value (Vafa et al., 2025). Even at the scale of component-level machine learning models, imposing physical constraints has not consistently improved accuracy or stability (Kelp et al., 2022; Sturm et al., 2023). Most physics-informed models do not satisfy the physics by construct and instead rely on soft constraints that need to be probed across a pre-defined range of physics. Therefore, the generalization challenge can still persist if the model has not been probed across all of the desired physics. There is little evidence to expect a different outcome at the foundation model scale, where the number of model transformations that must be constrained is far larger. Post-hoc validation avoids this tradeoff. Validating a model's outputs (and internals) against physical laws after training checks physical consistency without these restrictions.

## 4. Recommended Activities for 2028-2037

We recommend three activities for the coming decade. Each supports the physical evaluations described above and lies within the reach of coordinated investment rather than individual studies.

**Open, AI-ready datasets for physical evaluation.** Physical evaluation requires data against which an AI model's behavior can be checked. This data does not yet exist in AI-ready formats, but it can be assembled. Current benchmarks compare forecasts to reanalysis using skill metrics such as RMSE and correlation, and while some now include case studies of extreme events (Huang et al., 2026), none provide the detailed process-level information needed to determine whether a model's behavior is physically consistent (Dueben et al., 2022). Datasets built for this purpose would pair observed and modeled fields with the physical quantities that govern them, which vary by domain and might include emissions and reaction rates for atmospheric chemistry or heat and carbon fluxes for the land surface. They would include the record-breaking and compound events under which AI models are most likely to fail and would extend beyond weather to the ocean and land domains to which these models are increasingly adapted. Standardized, AI-ready benchmarks exist for weather skill but not for physical evaluation in other domains, such as the sparser and more heterogeneous observations of most other Earth system processes. Assembling this data in AI-ready, cloud-native formats such as Zarr and Icechunk is a scientific priority (Kelp et al., 2025). Because archiving every relevant field in full is impractical and data archives already ask modeling studies to reduce what they store, NASA could complement curated datasets with cloud-deployable model containers that regenerate additional physical diagnostics on demand. NASA is well positioned to lead this effort by curating and labeling existing archives, reanalysis products, and model output the agency already holds.

**A new standard for reporting physical evaluation.** The coming decade should establish a new, shared standard for how the physical behavior of an AI model is reported. The performance of these models should be characterized consistently rather than left to the discretion of each developer. Such reporting differs markedly from benchmark scores. A benchmark skill metric records how closely a forecast matches a reference, whereas physical evaluation reporting records whether the model behaves in accordance with physical laws. For example, a model release would document that the model conserves mass to within a stated tolerance, that a prescribed change in emissions produces the expected change in a pollutant, and that its skill was verified under both typical and extreme conditions. Such reporting could build on existing community evaluation frameworks (e.g., ESMValTool; Eyring et al., 2020) and should also specify the preprocessing applied to the training data, including normalization statistics, masking, and unit conventions. These details are rarely released, yet without them a model cannot be reliably evaluated or fine-tuned even on data from the instrument on which it was trained. Preprocessing should be considered part of the model, not simply an implementation detail. A prospective user could then determine whether a model is suited to a given application rather than inferring reliability from headline accuracy. This extends the documentation and quality standards NASA already maintains for its observational products to AI models and the products derived from them. Adopting such a standard while the field is small is far more tractable than imposing one after these models have spread across agencies, companies, and applications. It would also make the representation of physical processes, rather than forecast skill alone, the shared measure of whether a model can be trusted.

**A research program in the safety of Earth science AI models.** The evaluation methods in Section 3 are not yet common practice for Earth system AI models and establishing them is a central scientific task for the coming decade. Chief among them is mechanistic interpretability, which examines a model's internal features and computations to determine how it produces a forecast rather than accepting an AI prediction at face value. A research program in this area would establish how pretraining and fine-tuning shape a model's physical behavior, whether its internal representations correspond to physical processes, and how a model behaves under conditions it never encountered in training. Automated and agentic evaluation, in which AI systems assist in probing and interpreting other models, may be a promising direction as these methods continue to scale. This work parallels the interpretability research now common to the safety of large language models. The case for Earth science foundation models may be even stronger. A language model can be prompted to expose intermediate reasoning (i.e., “chain-of-thought reasoning”), whereas an Earth foundation model offers no such window, so interpreting its internal representations is the only means of determining how it arrives at a forecast. This research would let the community anticipate the conditions under which a model will fail before it is deployed rather than discovering them afterward. Because a foundation model builds its own internal representation of the Earth system, interpreting what it has learned can also expose relationships that are absent or misrepresented in conventional models, turning the evaluation of AI into a source of physical insight. Realizing this will require a scientific workforce fluent in both Earth science and the evaluation of AI systems, along with the governance capacity to act on what that evaluation reveals. The field does not yet cultivate this combination and must deliberately build it.

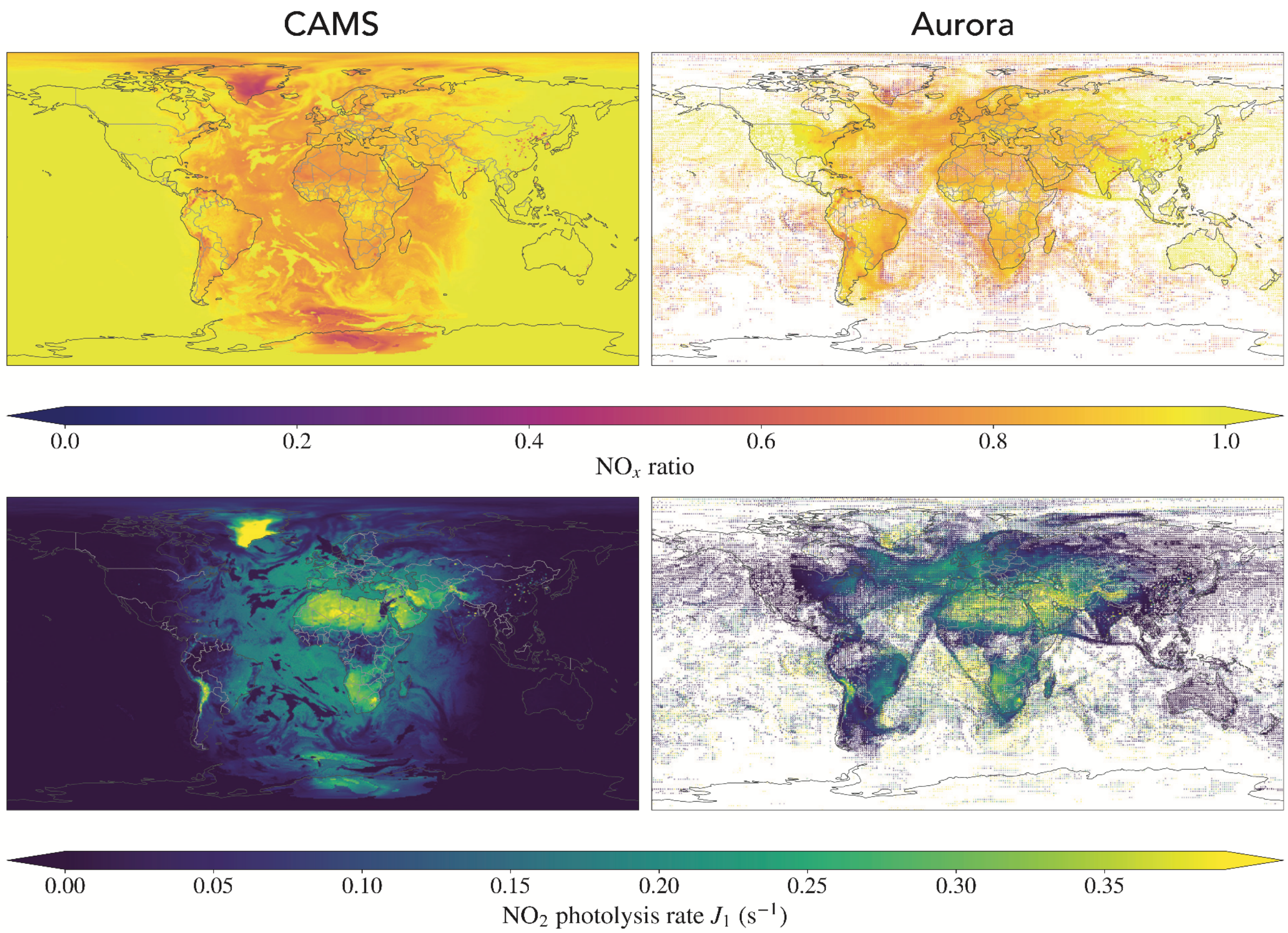

**Figure 1.** Calculated $NO_2$ photolysis rate and $NO_x$ ratio in CAMS and the Aurora Earth system foundation model fine-tuned for atmospheric chemistry. Both simulations are on September 1, 2022 at 12:00 UTC. White areas on the map indicate where the model predicts a negative concentration for at least one component species within the calculation. Reproduced from Hu et al. (2026).

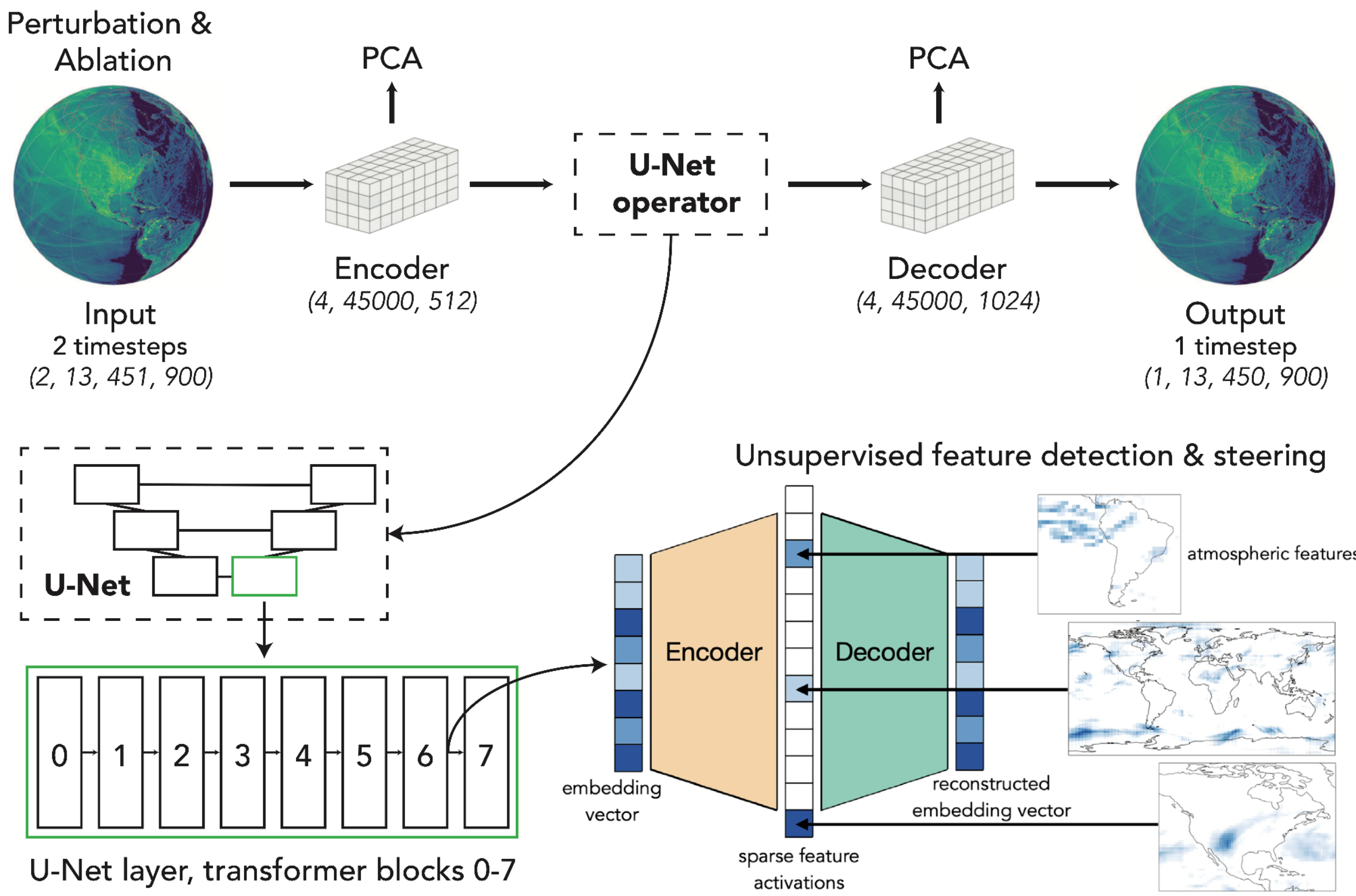

**Figure 2.** A framework for the physical evaluation of AI models in Earth science, illustrated for a foundation model. Three approaches probe how a model produces its forecasts. Inputs are perturbed and the output response analyzed. Latent representations are examined with principal components analysis (PCA). Sparse autoencoders are trained on each transformer block to discover interpretable features. These methods generalize across AI models. Reproduced from Hu et al. (2026) applied to the Aurora Earth system foundation model.